\documentstyle[twoside,fleqn,rgs_workshop,psfig]{article}
\newcommand{\etal}{{\it et al.} }
\newcommand{\asca}{{\it ASCA} }

\newcommand{\ginga}{{\it Ginga} }

\newcommand{\rosat}{{\it ROSAT} }
\newcommand{\xmm}{{\it XMM-Newton} }
\newcommand{\chandra}{{\it Chandra} }

\begin{document}

\title{The Starburst-AGN of NGC\,1808 Observed with \xmm \thanks{Based on observations obtained with {\it XMM-Newton}, an ESA science
mission with instruments and contributions directly funded by ESA Member
States and NASA}
}

\author{E. Jim\'enez Bail\'on
\address{LAEFF-INTA, POB 50727, E-28080 Madrid, Spain}
M. Santos-Lle\'o 
\address{ \xmm Science Operations Centre, VILSPA,
ESA, POB 50727, E-28080 Madrid,Spain}
M. Dahlem
\address{ESO, Alonso de Cordova 3107, Vitacura, Casilla 19001,
Santiago 19, Chile}
M. Ehle $^{b}$  M. Guainazzi $^{b}$
J. M. Mas Hesse $^a$
\address{Centro de Astrobiolog\'{\i}a (CSIC-INTA), E-28850 Torrej\'on
de Ardoz, Madrid, Spain} 
}

\begin{abstract}

NGC\,1808 is  a nearby spiral galaxy that  harbours an  active central
region with an extent of $20''$ ($\approx 1$ kpc).
Previous X-ray  and optical/NIR  observations have provided convincing
evidence for the existence of a starburst and an  AGN. We present here
preliminary  results  of the  analysis  of \xmm  data. We  show a weak
high-resolution  soft X-ray spectrum  with only emission lines typical
of a starburst.  Our analysis of  the {\it EPIC-pn} spectrum shows two
thermal components, but there  is an additional,  hard X-ray power law
tail that is most likely due to  an obscured active nucleus. Thus, our
data show for  the  first time the   presence  of emission  from  both
components, AGN and starburst, in one observation.

\end{abstract}

\maketitle

\section{Introduction}

NGC\,1808, classified as a Sbc pec galaxy \cite{sandage} is located at
a distance  of  10.9 Mpc (H$_0$=75   kms\,$^{-1}$Mpc$^{-1}$, 1$''$= 53
pc). Images in different wavebands suggest a high star-formation level
in  the  central region with  $\sim  20''$(i.e., a diameter of $\approx 1$ kpc): there are
several optical  hot spots associated  with  HII regions, luminous and
compact knots in radio and IR
\cite{saikia,kotilainen}, that do not coincide with  the optical hot spots
and probably are supernova  remnants (SNR) or complexes of   unresolved
SNRs; dust filaments explained as outflowing material driven by supernovae
\cite{heckman}.  A recent interaction of NGC\,1808 
with its companion NGC~1792  could explain both the 
intense star-formation activity   and the peculiar  morphology of  the
galaxy \cite{dahlem,koribalski}.

The nature of the nucleus is unclear: it is classified as Seyfert~2, using
the optical nuclear emission lines~\cite{veron}; it was classified as an
obscured Seyfert but also as a hidden starburst based on 
measurements of  polarized optical light  
\cite{scarrott}; the nucleus is claimed to have a strong non-stellar component 
because only 10\% of the IR radiation observed by {\it ISO} could be
interpreted as emission related to star formation~\cite{siebenmorgen}. 
In the X-ray band, Dahlem et al.~\cite{dahlem} and
Junkes~\etal~\cite{junkes} favour a stellar origin based on \rosat
observations (0.1--2.4 keV) but do not discard other hypotheses. 
Awaki and  Koyoma~\cite{awaki93} interpret the \ginga X-ray data (1.5-37 keV) 
with an obscured nucleus, however, Awaki et al. \cite{awaki96},
using \asca (2--10 keV) point out that the hard X-ray spectrum may be the
result of starburst activity and but the long-term variability from the
\ginga and \asca observations suggests a Seyfert nature of the nucleus. 

A \chandra image of the central region of NGC\,1808 shows that
the X-ray  emission is extended and consists of various circumnuclear
sources that coincide with ultraviolet (UV) HII regions as observed with the
{\it HST}~\cite{zezas}. 

The spectral analysis of the \xmm data shown here sheds  new  light on
the  nature  of the nuclear activity.  In this  work, we
present preliminary results  of  the   analysis  of the optical, UV and
X-ray \xmm observations.

\section{The Data: Observations and Images}
e
The \xmm observations were performed on April 6, 2002. The {\it EPIC-pn} 
exposure time was 36000 s, in extended full frame mode and with the thin
filter. The {\it RGS} time was 40000 s. {\it OM} was used with the U, UVW1
and UVW2 broad band 
filters, plus the two, optical and UV, grisms. The data has been processed
with the Science Analysis System, SAS, version 5.3.0 for  {\it EPIC-pn}
and{\it OM} and 5.3.3 for {\it RGS}. 

\begin{figure}[htb] 
{\psfig{file=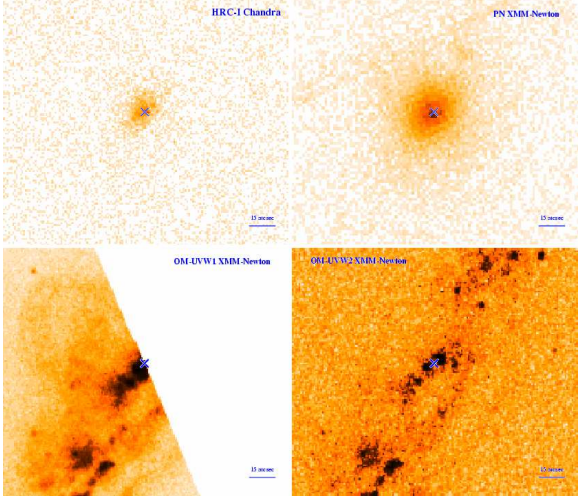,width=3.0in,angle=0}}
\caption{NGC\,1808 images. From left
to right and  from top  to bottom,  \chandra archive image; {\it EPIC-pn}
image; {\it OM} UVW1, 291 nm, image; and {\it OM} UVW2 (212 nm) image. 
Some data is lost in  the {\it OM} UVW1  image due to  technical 
problems. The location of of the {\it VLA} radio
nucleus is marked with a cross. 
}\label{uvx}
\end{figure}

We have compared the \chandra, {\it EPIC-pn} and {\it OM} UV images. As 
shown in Fig.~\ref{uvx}, the maximum emission in each of these images
coincides with the location of the {\it VLA} radio
nucleus~\cite{saikia}. The X-ray emission is extended and elongated in the
direction of the UV structure.  

\begin{figure}[htb] 
{\psfig{file=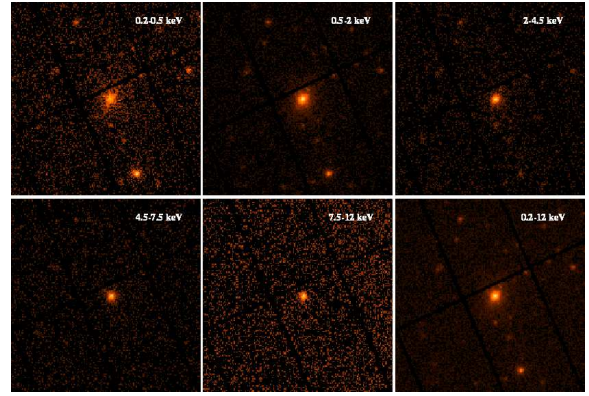,width=3.0in,angle=0}}
\caption{{\it EPIC-pn}  images: 
from left to right and from top to bottom, 0.2-0.5 keV, 0.5-2 keV, 2-4.5
keV, 4.5-7.5 keV,  7.5-12 keV and  0.2-12 keV. 
 }\label{xbands}
\end{figure}

Fig.~\ref{xbands}   shows {\it EPIC-pn}  images of   NGC\,1808 in several
energy bands.  As  the energy  increases,  the nuclear emission becomes more
focused and the non-nuclear sources disappear. 

\section{Spectral Analysis of the {\it EPIC-pn} Data}

{\it  EPIC-pn} data have been  used to perform  a spectral analysis of
NGC\,1808  in the 0.15-15  keV energy band  with moderate resolution 
(80\,eV at 1\,keV) and using single and double events. 

\begin{figure}[htb] 
{\psfig{file=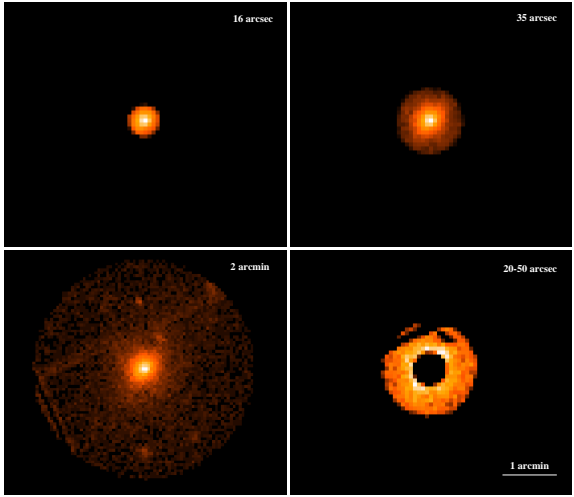,width=3.0in,angle=-90}}
\caption{{\it EPIC-pn} images of the regions considered for the spectral
analysis. The scale is the same for all images and indicated in the
lower-right frame. The extraction radii are given in the top right corner
of each frame.  
 }\label{pn_regions}
\end{figure}

\begin{figure}[htb] 
{\psfig{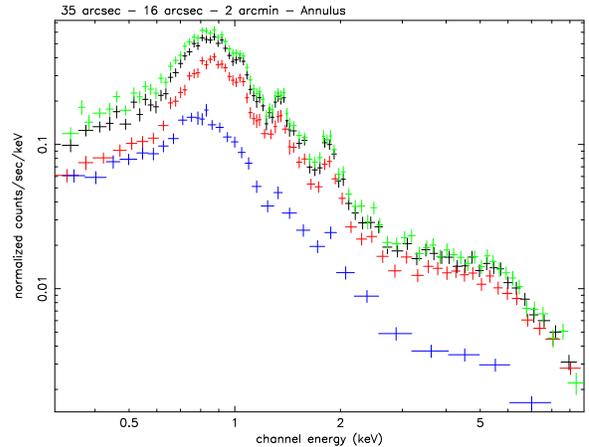}}
\caption{{\it EPIC-pn} spectra of the three  circular regions and the
annulus shown in Fig.~\ref{pn_regions}. Red, black, green and blue
correspond to radii of 16$''$, 
35$''$   and 2$'$ and 20$''$-50$''$ annulus, respectively.}
\label{pn_spectra}
\end{figure}

In order to search for differences in the X-ray spectrum of the nucleus 
and its surrounding HII regions, we have considered three circular regions
and one annulus. The circles have radii of 16$''$  (850\,pc), 35$''$
(1.9\,kpc) and 2$'$  (6.4\,kpc) and the annulus has external and 
internal radii of 20$''$ and 50$''$ (1-2.7 kpc) and also excluding 
an extra-nuclear  source. 
Figures~\ref{pn_regions} and~\ref{pn_spectra} show the extraction regions
and the spectra, respectively. 

The  spectra  of the  35$''$ and 2$'$ regions are very similar both in
shape and intensity.  The spectrum of the 16$''$ region, although in 
good agreement with the two previous ones above 2 keV, is clearly weaker in
the soft band.  

\subsection{The {\it EPIC-pn} Nuclear Spectrum}

\begin{figure}[htb] 
{\psfig{file=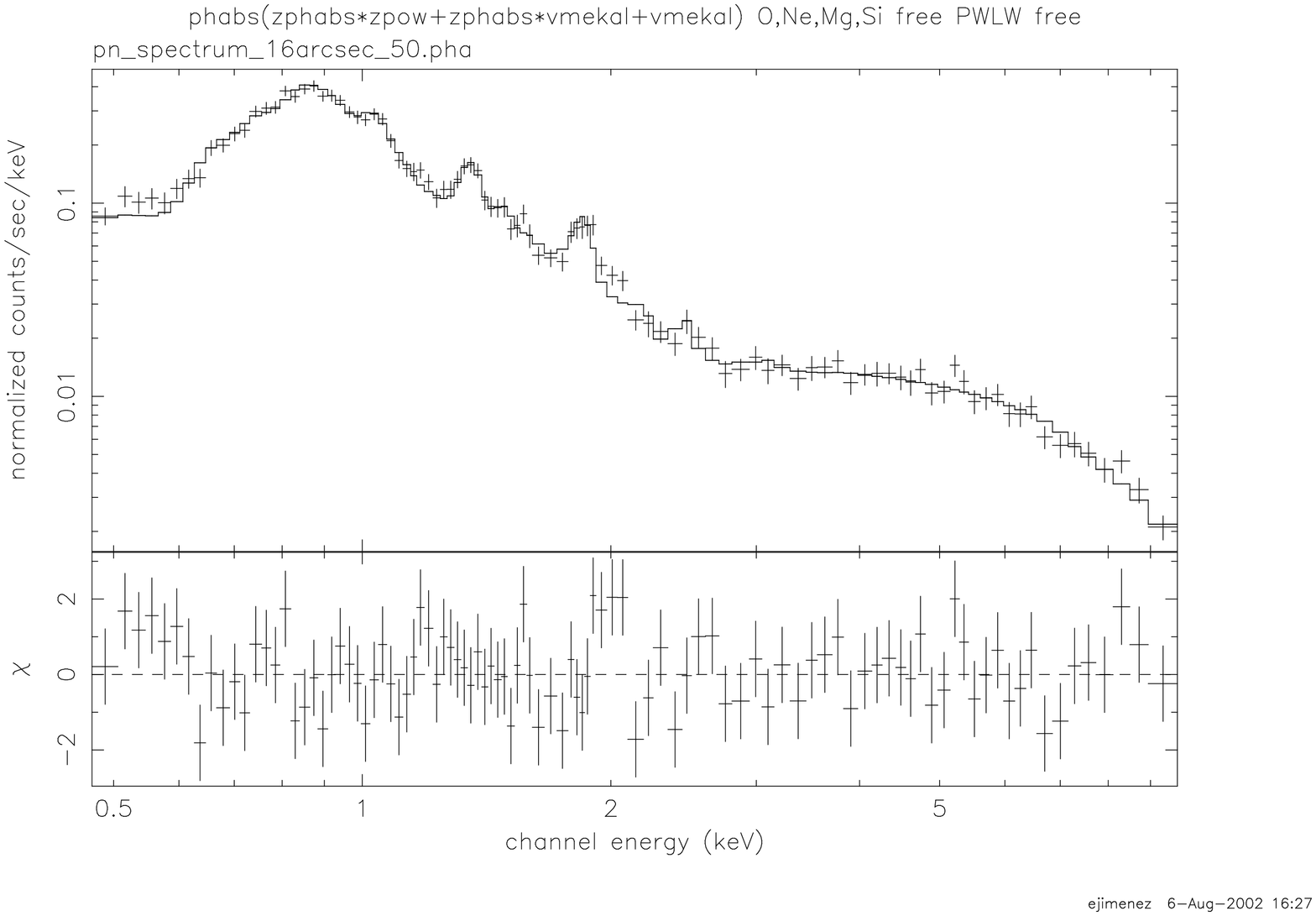,width=3.0in,angle=-90}}
\caption{{\it EPIC-pn} spectrum, best fit model and residuals of the
r=16$''$ nuclear region. }\label{pn_nucleus}
\end{figure}

The {\it  EPIC-pn} spectrum of the  NGC\,1808 nucleus, r=16$''$, 
shows prominent  Si and Mg emission lines.
Fig.~\ref{pn_nucleus} shows the
0.5--10\,keV spectrum, the best fit model, $\chi_{\nu}^2$=1.04 for 88
dof, and the residuals.  
The  model  includes, apart  from  the galactic  absorption due to the
Galaxy with N$_H=3.23\times10^{20}$\,cm$^{-2}$,  a power law,   $\Gamma =
1.350^{+0.22}_{-0.29}$, absorbed by  an extra 
Hydrogen column of N$_H$=5.4$^{+2.1}_{-1.6}\times10^{22}$\,cm$^{-2}$ and
two thermal components with temperatures of kT=0.53$^{+0.07}_{-0.05}$
keV   and  kT=0.62$^{+0.02}_{-0.04}$  keV,   the former also absorbed
by an extra Hydrogen column of 
N$_H$=1.3$^{+0.3}_{-0.2}\times10^{22}\,$cm$^{-2}$. The abundances of  O,  
Ne,  Mg  and Si, let free in the fits, are 1.6$^{+0.4}_{-0.3}$,
1.9$^{+0.6}_{-0.7}$, 1.4$^{+0.4}_{-0.3}$ and 0.7$^{+0.2}_{-0.2}$  times
the solar value, respectively. No Fe~K$\alpha$ line is included. 

The r=16$''$ region luminosities corrected  for 
absorption are shown  in Table~\ref{lumin}. The thermal  components
dominate in the soft X-ray band while the power law is dominant in the
2-10\,keV band. Table~\ref{lumin} also shows 
the luminosity of a r=62$.\!\!''$5 region measured with {\it EPIC-pn} and
with \rosat \cite{junkes}.  The large discrepancy  could   be
due to  a  long-term variability over several  years, from February   1991 to
April 2002, already  suggested by Awaki~\etal~\cite{awaki96} when comparing 
\ginga and  \asca data.  Short-term variability  has not been  detected
during the \xmm observation.

\begin{table}[htb]   
\caption{\centerline{NGC\,1808 unabsorbed X-ray luminosities}}
\label{lumin} 
\begin{tabular}{lll}
\hline
{\bf Region} &  L$_{0.1-2.4\,keV}$ & L$_{2-10\,keV}$  \\
radius & $10^{39}$\,erg\,s$^{-1}$ & $10^{39}\,$erg\,s$^{-1}$ \\
\hline
16$''$ &9.5  & 15 \\
\,\,{\it Power law} & 0.3 ($\sim 5\%$) & 14 ($\sim 90\%$)\\
\,\,{\it Thermal }  & 9.2 ($\sim 95\%$) & 1.2 ($\sim 10\%$)\\
\hline
62$.\!\!''$5 & 12 & 15.5 \\ \hline
62$.\!\!''$5 \rosat & 140 & - \\
\hline

\hline 
\end{tabular}
\end{table}

\section{Spectral Analysis of the High Resolution {\it RGS} Data}

\begin{figure}[htb] 
\vspace{5pt}
{\psfig{file=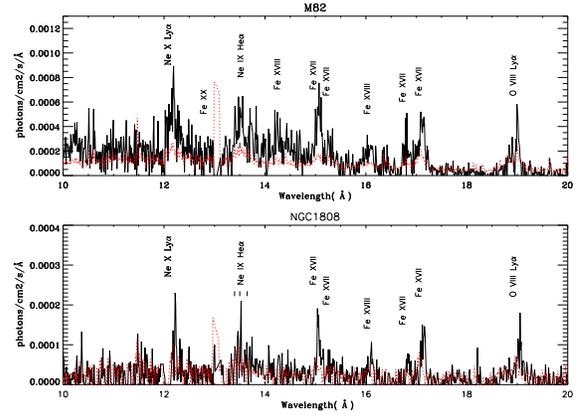,width=3.0in,angle=90}}
\caption{{\it RGS} spectra of M82 and NGC\,1808, upper and lower
panel, respectively. The  errors are shown in red, dotted line. The
identification of M82 lines is from Read \& Stevens\,\cite{read}.}\label{rgs} 
\end{figure}

Fig.~\ref{rgs} compares the combined {\it RGS1} and {\it RGS2} 
spectrum of NGC\,1808 with that of M82, a proto-typical starburst galaxy. 
Both spectra have been generated with the {\tt rgsfluxer} SAS 5.3.3 task. 
The {\it RGS} spectrum of NGC\,1808 shows no continuum emission 
above the noise level, but it does exhibit emission lines. 
Fig.~\ref{rgs} shows that these 
have wavelengths and relative intensity ratios very similar to the
strongest lines identified in M82 \cite{read}. They correspond to 
Ly$\alpha$ emission lines from Ne X and O VIII, transitions of He-like
Ne IX ions, and Fe-L  emission lines from Fe XVII and 
Fe XVIII. The weaker lines visible in the spectrum of M82 are not detected
in NGC\,1808, because of the lower signal-to-noise ratio. 
The similarity in the line ratios suggests that there
is a common origin of the soft X-ray emission in both galaxies, i.e. 
thermal emission from a hot and extended gas component as shown for M82
\cite{read}.  This  result confirms the detection of the starburst in
NGC\,1808 which dominates the soft X-ray band emission.

\section{Summary and Conclusions}

The \xmm EPIC images show extended X-ray emission in agreement with
previous HRI-\chandra results.   There is a  correlation between X-ray
and  ultraviolet emission:   the location  of  the maximum  luminosity
corresponds to  an unresolved point-like source  and coincides in both
spectral ranges. Comparing {\it EPIC-pn} spectra of several regions in
the central part of NGC\,1808,  it is inferred that the bulk of the
emission originates from the nucleus, although in the soft X-ray band
the contribution of the  circumnuclear regions is  not negligible.  The
{\it EPIC-pn} spectrum  of the nucleus (r=16$''$, 850 pc)  is 
explained by two thermal components which account for the main part of
the  soft  X-ray emission, but  an additional  power  law component is
required to explain the hard X-rays.  The luminosity of the unresolved
nucleus in the  0.2--10 keV band is $\sim\,2\times10^{40}\,$erg\,s$^{-1}$.
The {\it EPIC-pn} result suggests the presence of an additional obscured
AGN-like component. 
The data obtained with {\it RGS} exhibit emission  lines similar in wavelength
and relative intensity ratios to the ones found for the prototypical
starburst galaxy M82. This result confirms the detection of a nuclear
starburst in NGC\,1808 which dominates the total emission spectrum including
the continuum in the soft X-ray regime.


\normalsize



\small
\begin{thebibliography}{9}

\bibitem{sandage} Sandage, A.~\& Tammann, G.~A.\ 1987, Carnegie Institution of Washington Publication, Washington: Carnegie Institution, 1987, 2nd ed.

\bibitem{saikia} Saikia, D.~J., Unger, S.~W.,
Pedlar,  \etal \ 1990, MNRAS, 245, 397 

\bibitem{kotilainen} Kotilainen, J.~K., 
Forbes, D.~A., Moorwood, A.~F.~M., van der Werf, P.~P., \& Ward, M.~J.\ 
1996, A\&A, 313, 771 

\bibitem{heckman} Heckman, T.~M., Armus, L., \& Miley, G.~K.\ 1990, APJS, 74, 833 

\bibitem{dahlem} Dahlem, M., Bomans, D.~J., \& Will, J.\ 1994, ApJ, 432, 590 

\bibitem{koribalski}  Koribalski, B., Dahlem, M., Mebold, U., \& Brinks, E.\ 1993, A\&A, 268, 14 

\bibitem{veron} V\'eron-Cetty, M.-P.~\& V\'eron, P.\ 1985, A\&A, 145, 425 

\bibitem{scarrott} Scarrott, S.~M., Draper,
P.~W., Stockdale \etal \ 1993, MNRAS, 264, L7 

\bibitem{siebenmorgen}Siebenmorgen, R., Kr{\" u}gel, E., \& Laureijs, R.~J.\ 2001, A\&A, 377, 735 

\bibitem{junkes}Junkes, N., Zinnecker, H., Hensler \etal \ 1995, A\&A, 294, 8 

\bibitem{awaki93} Awaki, H.~\& Koyama, 
K.\ 1993, Advances in Space Research, 13, 221 

\bibitem{awaki96} Awaki, H., Ueno, S., Koyama \etal
\ 1996, PASJ, 48, 409 

\bibitem{zezas} Zezas A.,Ward M., Fabbiano G. \etal \ 2002{\footnote {http://hea-www.harvard.edu/HEAD-preprints/2000/November/zezas/paper.ps}}
 
\bibitem{read}Read, A.~M.~\& Stevens, I.~R.\ 2002, MNRAS, 335, L36 







\end{thebibliography}
\end{document}